\documentclass[aps,pra,float,twocolumn,superscriptaddress]{revtex4}
\usepackage{graphicx}
\usepackage{braket}
\usepackage{amsmath}
\usepackage[colorlinks, citecolor=blue, urlcolor=red, linkcolor=blue]{hyperref}
\usepackage{amsfonts}
\usepackage{float}
\usepackage[font=small,format=plain,labelsep=period,labelfont=normal,textfont=normal,singlelinecheck=false,justification=raggedright]{caption}
\usepackage{subcaption}
\usepackage{xcolor}
\usepackage{comment}
\usepackage{textcomp}
\usepackage{amsthm}
\usepackage{amssymb}
\usepackage{physics}
\usepackage{mathtools}

\begin{document}
%======================================================================
	\title{Estimation of one-dimensional discrete-time quantum walk parameters by using machine learning algorithms}

	\author{Parth Rajauria}
	%\email{}
	\affiliation{The Institute of Mathematical Sciences, C. I. T. Campus, Taramani, Chennai 600113, India}
	\affiliation{Indian Institute of Science Education and Research(IISER) Tirupati, Tirupati, 517507, India}
	\author{Prateek Chawla}
	\affiliation{The Institute of Mathematical Sciences, C. I. T. Campus, Taramani, Chennai 600113, India}
	\affiliation{Homi Bhabha National Institute, Training School Complex, Anushakti
Nagar, Mumbai 400094, India}
	\author{C. M. Chandrashekar}
	\email{chandru@imsc.res.in}
	\affiliation{The Institute of Mathematical Sciences, C. I. T. Campus, Taramani, Chennai 600113, India}
	\affiliation{Homi Bhabha National Institute, Training School Complex, Anushakti
Nagar, Mumbai 400094, India}
%======================================================================
%======================================================================	

\begin{abstract}
Estimation of the coin parameter(s) is an important part of the problem of implementing more robust schemes for quantum simulation using quantum walks. 
We present the estimation of the quantum coin parameter used for one-dimensional discrete-time quantum walk evolution using machine learning algorithms on their probability distributions. We show that the models we have implemented are able to estimate these evolution parameters to a good accuracy level. We also implement a deep learning model that is able to predict multiple parameters simultaneously. Since discrete-time quantum walks can be used as quantum simulators, these models become important when extrapolating the quantum walk parameters from the probability distributions of the quantum system that is being simulated.
\end{abstract}	
	
%======================================================================	
	
\maketitle

%======================================================================
\section{Introduction}
\label{sec:intro}
%======================================================================

Quantum walks \cite{R58, F86} are the quantum counterparts of classical random walks, and are therefore used as the basis to generate relevant models for controlled quantum dynamics of a particle. Much like a classical random walk, the formalism for quantum walks has also developed in two forms - the discrete-time quantum walk (DTQW) and the continuous-time quantum walk (CTQW). Both the formalisms exhibit features that enable them to effectively realise quantum computational tasks \cite{DW08, C09, A07,MSS07,BS06,FGG07,K08}. Quantum superposition and interference have the effect of allowing the quantum walker to have a quadratically faster spread in position space in comparison to a classical walker \cite{P88, ADZ93, M96,K03, CCD03, CG04, VA12}. This has applications in modelling dynamics of several quantum systems, like quantum percolation \cite{CB14,KKNJ12,CAC19}, energy transport in photosynthesis \cite{ECR07, MRLA08}, graph isomorphism \cite{DW08}, quantum algorithms \cite{CMC20, PMCM13}, and even in generating a scheme for universal quantum computation \cite{SCSC19, SCASC20}.

Much like a classical walk, the dynamics of a walker undergoing CTQW can be described only by a position Hilbert space, whereas a walker performing DTQW requires an additional degree of freedom to express its controlled dynamics. This is realised by the coin Hilbert space, which is an internal state of the walker, and provides the relevant additional degree of freedom for the walker. Tuning the parameters and evolution operators of a DTQW enables the walker to simulate several quantum mechanical phenomena, such as topological phases \cite{COB15, SRFL08, KRBD10, AO13}, relativistic quantum dynamics \cite{S06,CBS10, MC16,C13,MBD13,MBD14,AFF16,P16,RLBL05}, localization \cite{J12, C12, CB15}, and neutrino oscillations \cite{MMC17,MP16}. Quantum walks have been experimentally implemented in various physical systems, such as photonic systems \cite{SCP10,BFL10,P10}, nuclear magnetic resonance systems \cite{RLBL05}, trapped ion systems \cite{SRS09,ZKG10}, and cold atoms \cite{KFCSWMW09}. 

It is known that the coin parameters of a quantum walk play a significant role in determining the overall dynamics of the system. For instance, the coin parameters of a split-step quantum walk determine topological phases \cite{COB15,SRFL08,KRBD10}, neutrino oscillation \cite{MMC17,MP16}, and the mass of a Dirac particle \cite{MC16}. Thus the coin parameter is an important piece of the puzzle while using a quantum walk as a quantum simulation scheme. It thus becomes a crucial problem to be able to estimate the parameters of a quantum walk in order to facilitate better quantum simulations and also for further research into modelling realistic quantum dynamics. 

In this regard, the problem partially becomes one of finding a pattern hidden in this complex data, and an effective approach is to use an algorithm that automates its learning process. In this context, the term 'learning' aptly described in \cite{Mitchell97} : A computer program is said to learn from experience E with respect to some class of tasks T and performance measure P, if its performance at tasks in T, as measured by P, improves with experience E. The task T in this particular case is defined as, to output a function $f(\mathbb{R}^N \rightarrow \mathbb{R})$, such that the input vector corresponds to the known probability distribution after $N$ steps and the output corresponds to the parameter $\theta$. It may, however, be noted that due to the No Free Lunch theorem \cite{W96}, the machine learning strategies for different types of quantum walks are not likely to be the same as for this particular case. 

It is known that the Quantum Fisher Information (QFI) \cite{P11,TP13,S13, TA14} $H_w(\theta)$ of a quantum walker's position quantifies a bound to the amount of information that can be extracted from the probability distribution \cite{P12,SCP19}. It has also been shown that it is indeed possible to estimate the coin parameter of a DTQW with a single-parameter coin. However, the approaches that rely on QFI to predict the parameters in a DTQW share a similar constraint, viz, even though $H_w(\theta)$ is a continuous, single-valued function over $\theta$, the variations in the plot are small, and thus contribute to the error in determining $\theta$ \cite{SCP19}. In this work, we train various machine learning models and demonstrate that such models can indeed estimate the quantum walk parameters to a good accuracy. We also estimate the coin parameter and the number of steps for a DTQW simultaneously with a multilayer perceptron model, and demonstrate that it performs much better than a baseline model that does not learn with experience.

This paper is structured as follows. In section \ref{sec:dtqw} we introduce a standard DTQW, an SSQW, and describe the evolution operators that determine their dynamics. We give a brief overview of the machine learning models used and their parameters in Section \ref{sec:ml}. Section \ref{sec:res} details our results of training the machine learning algorithms and their performance. We wrap up the paper in section \ref{sec:conc} with a small discussion on our results and conclusions drawn.

%======================================================================
\section{Discrete-time quantum walk}
\label{sec:dtqw}
%======================================================================

The evolution of a walker executing DTQW is defined in a Hilbert space $\mathcal{H} = \mathcal{H}_C \otimes \mathcal{H}_P$, where $\mathcal{H}_C$ and $\mathcal{H}_P$ are the walker's coin and position Hilbert spaces, respectively. The coin Hilbert space has the basis states $\left\{ \ket{\uparrow}, \ket{\downarrow} \right\}$, and the position Hilbert space is defined by the basis states $\ket{x}$, where $x \in \mathbb{Z}$. The evolution is described by two unitary operators, known as the coin and shift operations. The coin operator affects the coin Hilbert space, and the shift operator evolves the walker in a superposition of position states, the amplitudes of which are determined by the coin operation. In case of the one-dimensional DTQW, the most general unitary coin operator is an $SU(2)$ matrix, which has three independent parameters. That said, even one- and two-parameter coins are very useful while simulating various systems. As an example, the single-parameter split step DTQW is very effective to simulate neutrino oscillations \cite{MMC17}, topological phases \cite{COB15}, and the Dirac equation \cite{MC16}. 
% TODO: introduction worthy
%It is thus crucial to obtain an understanding of how the parameters affect the dynamics of the quantum walk, in order to be able to tailor various quantum walks to model dynamics of quantum systems in a more realistic manner.

In a one-dimensional DTQW, the coin operation is an $SU(2)$ matrix, defined as, 
\begin{equation}
	\label{eq:eq2.1}
	\hat{C}(\theta, \xi, \zeta)
	= \begin{bmatrix} 
	e^{i\xi}\cos(\theta) & e^{i\zeta}\sin(\theta) \\
	-e^{-i\zeta}sin(\theta) & e^{-i\xi}cos(\theta) \end{bmatrix} \otimes \sum_{x\in \mathbb{Z}} \ket{x} \bra{x}.
\end{equation}

The shift operation is defined as,
\begin{equation}
	\label{eq:eq2.2}
	\hat{S} = \sum_{x} \bigg[\ket{\uparrow}\bra{\uparrow} \otimes \ket{x-1}\bra{x} + \ket{\downarrow}\bra{\downarrow} \otimes \ket{x+1}\bra{x}\bigg].
\end{equation}
The initial internal state of the walker is defined as,
\begin{equation}
	\label{eq:eq2.3}
	\ket{\psi_0} = \left(\alpha \ket{\uparrow} + \beta \ket{\downarrow}\right) \otimes \ket{x=0}
\end{equation}
and the complete evolution equation will be in the form, 
\begin{equation}
	\label{eq:eq2.4}
	\ket{\psi_N} = \big[ \hat{S}\hat{C} \big]^N \ket{\psi_0} = \hat{W}^N \ket{\psi_0},
\end{equation}
\noindent
where $N$ is the number of steps taken by the walker.

To obtain the the one-parameter coin form from Eq.~(\ref{eq:eq2.1}), we fix the values $\xi=0$ and $\zeta = -\frac{\pi}{2}$. This is the convention adopted in the remainder of this text. The one-parameter coin operator is thus given as,
\begin{equation}
	\label{eq:eq2.5}
	\hat{C}(\theta) = \begin{bmatrix}
	\cos(\theta) & -i\sin(\theta) \\
	-i\sin(\theta) & \cos(\theta)
	\end{bmatrix} \otimes \sum_{x\in \mathbb{Z}}\ket{x}\bra{x}.
\end{equation}

Fig.~\ref{fig:fig2.1} shows the probability distributions in position space of a walker performing a DTQW with different values of $\theta$ as in Eq.~(\ref{eq:eq2.5}).

\begin{table*}
	\begin{minipage}[c]{\textwidth}
		~
		%\begin{figure}[!ht]
		\centering
		\begin{tabular}{cc}
			\includegraphics[width=0.45\textwidth]{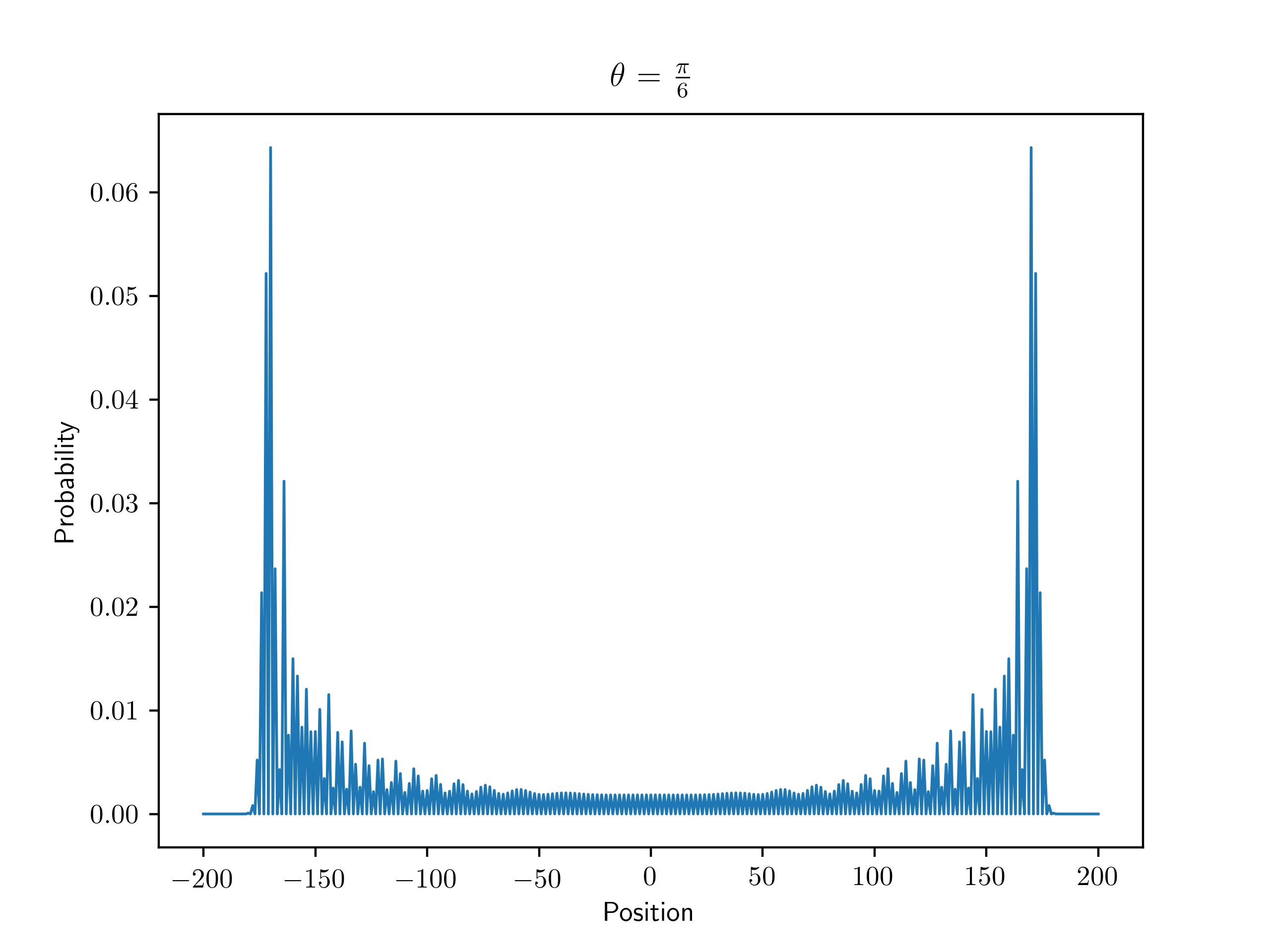} &
			\includegraphics[width=0.45\textwidth]{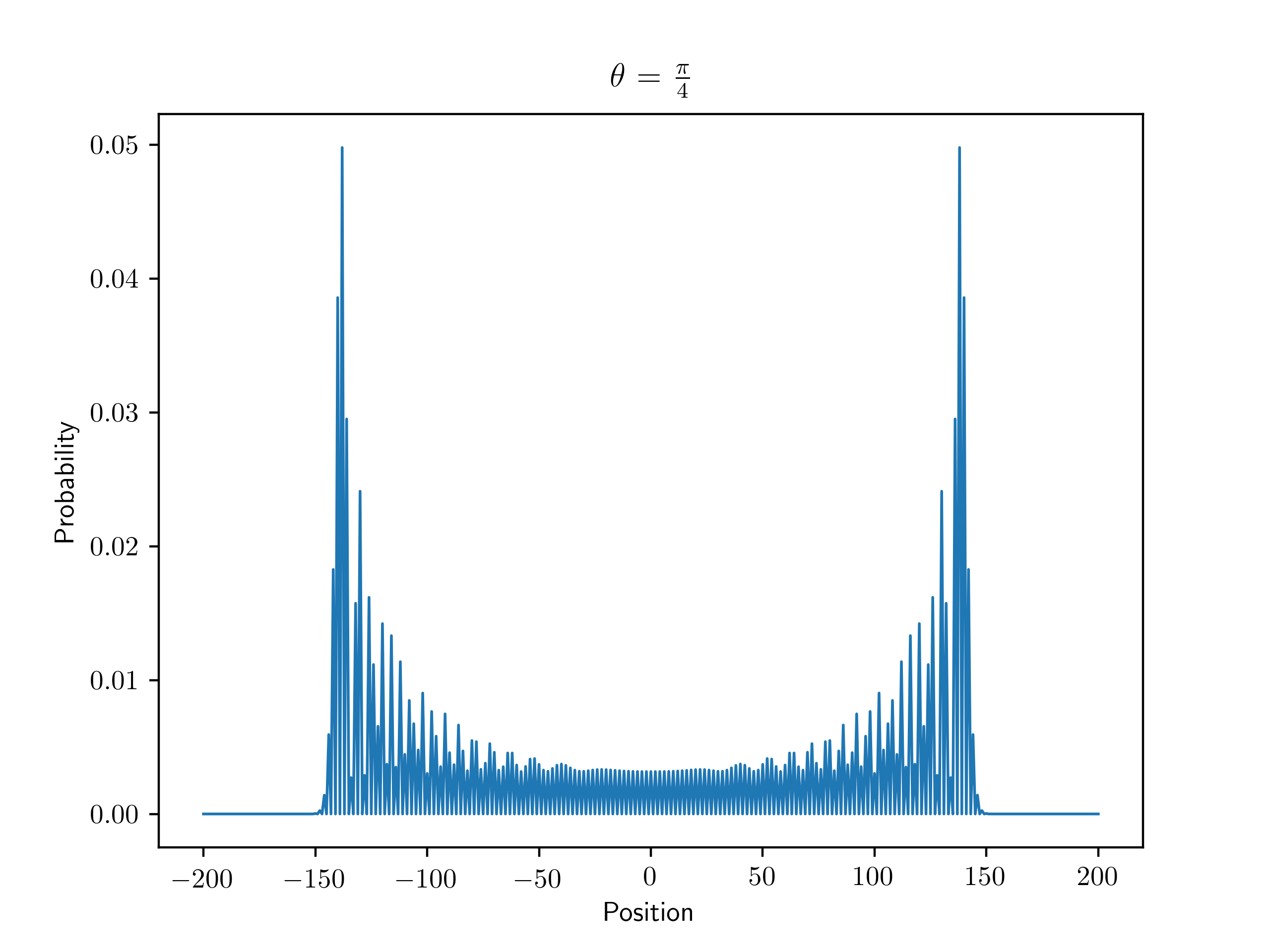} \\	
			(a) & (b)\\
			\includegraphics[width=0.45\textwidth]{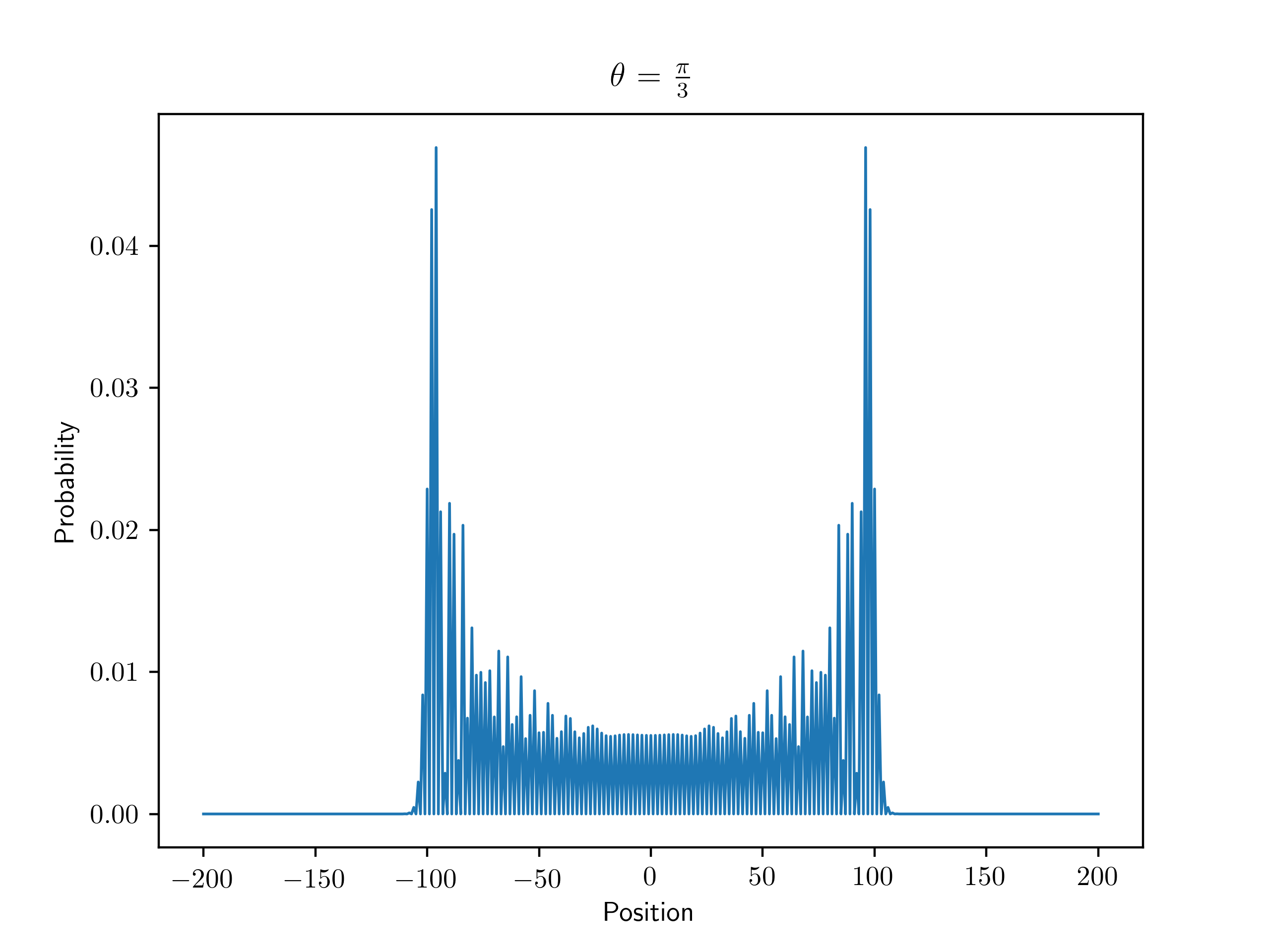} &
			\includegraphics[width=0.45\textwidth]{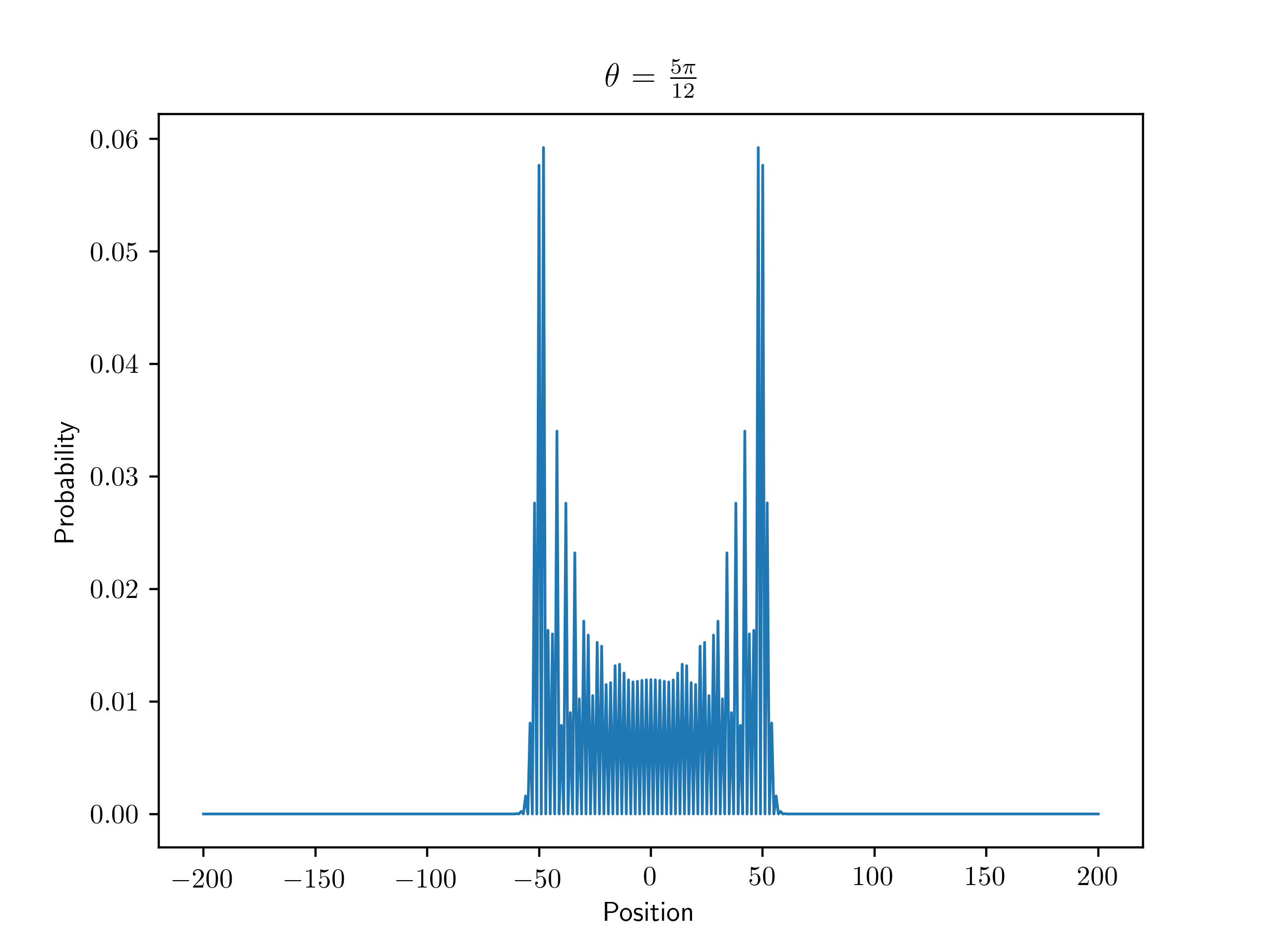} \\	
			(c) & (d)\\
		\end{tabular}
		\captionof{figure}{Position space probability distributions of a single walker performing a one-parameter DTQW in one dimension, obtained after 200 steps of walk. The figures (a), (b), (c) and (d) correspond to the cases where the parameter $\theta$ has been set to $\frac{\pi}{6}, \frac{\pi}{4}, \frac{\pi}{3}$, and $\frac{5\pi}{12}$, respectively. The distibutions are symmmetric as the initial state in each case was set to $\ket{\psi(0)} = \frac{1}{\sqrt{2}}(\ket{\uparrow} + i\ket{\downarrow})\otimes\ket{x=0}$, which is symmetric.
			\label{fig:fig2.1}}
	\end{minipage}
\end{table*}

A special case of a DTQW is the split-step quantum walk (SSQW), in which case the evolution operator $\hat{W}$ is given by a composition of two half-steps,
\begin{equation}
\label{eq:eq2.6}
	\hat{W} = \hat{S}_+\hat{C}_{\theta_2}\hat{S}_-\hat{C}_{\theta_1},
\end{equation}
\noindent
Where $\hat{C}_{\theta_j}$ represent the two coin operations, which are defined in the same form as in Eq.~(\ref{eq:eq2.5}), and $\hat{S}_\pm$ are directed shift operations, defined as,
\begin{equation}
	\label{eq:eq2.7}
	\begin{split}
		\hat{S}_+ &=  \sum_{x} \bigg[\ket{\uparrow}\bra{\uparrow} \otimes \ket{x}\bra{x} + \ket{\downarrow}\bra{\downarrow} \otimes \ket{x+1}\bra{x}\bigg] \\
		\hat{S}_- &= \sum_{x} \bigg[\ket{\uparrow}\bra{\uparrow} \otimes \ket{x-1}\bra{x} + \ket{\downarrow}\bra{\downarrow} \otimes \ket{x}\bra{x}\bigg]
	\end{split}.
\end{equation}
In a one-dimensional visualisation, an SSQW implements the case when the components that are in the direction of $\ket{\uparrow}$ experience a different coin ($\hat{C}_{\theta_1}$) than the components in the $\ket{\downarrow}$ direction, which experience the coin $\hat{C}_{\theta_2}$. The evolution of the walk is still in the form as shown in Eq.~(\ref{eq:eq2.4}), with the appropriate walk operator $\hat{W}$ substituted from Eq.~(\ref{eq:eq2.6}).

Fig.~\ref{fig:fig2.2} shows the probability distributions obtained by a walker executing a SSQW with multiple combinations of $\theta_1$ and $\theta_2$. Unlike the two peak probability distribution from the single coin DTQW, SSQW can results in  four peak probability distribution.
\begin{table*}
	\begin{minipage}[c]{\textwidth}
		~
		%\begin{figure}[!ht]
		\centering
		\begin{tabular}{cc}
			\includegraphics[width=0.45\textwidth]{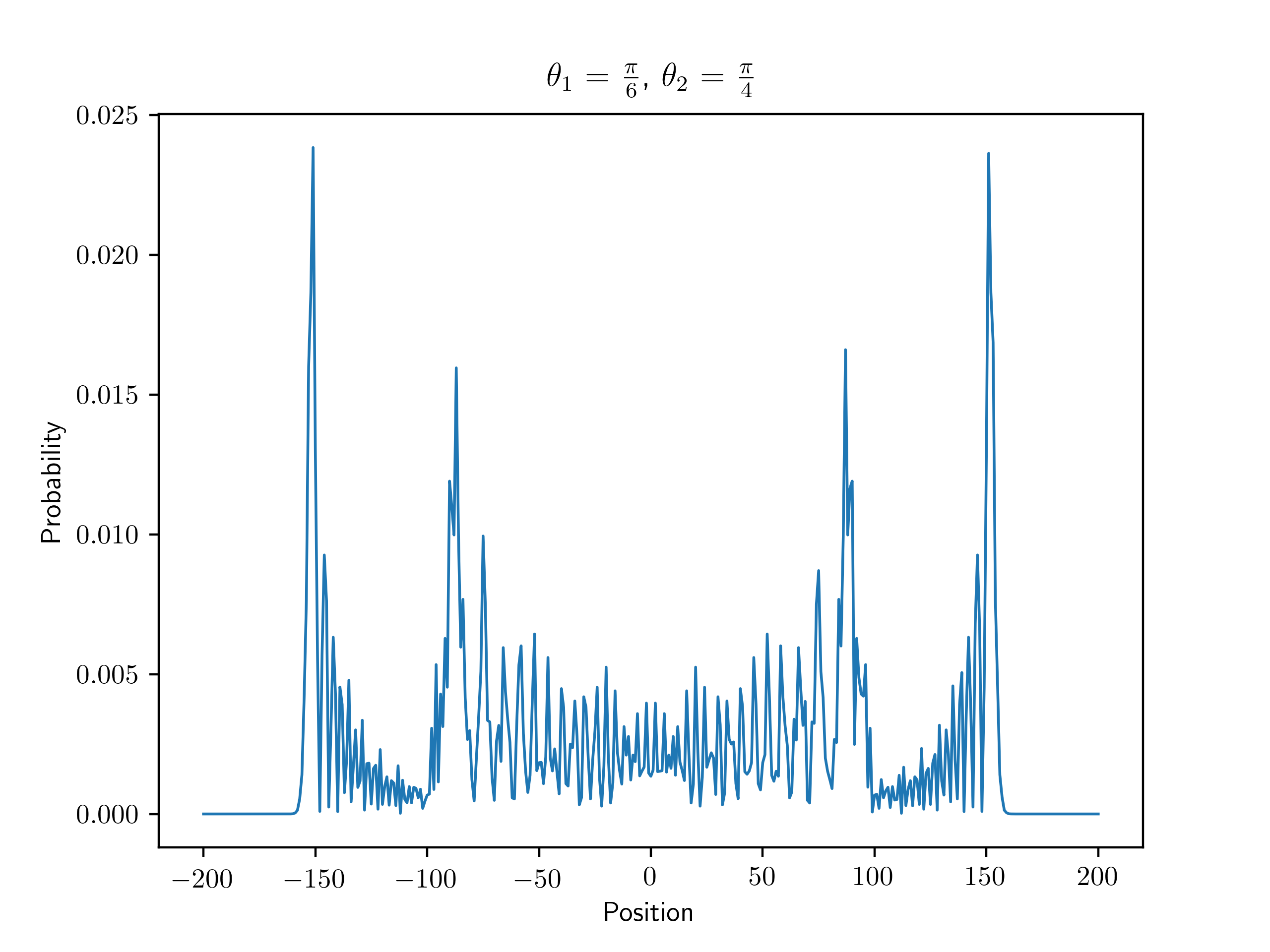} &
			\includegraphics[width=0.45\textwidth]{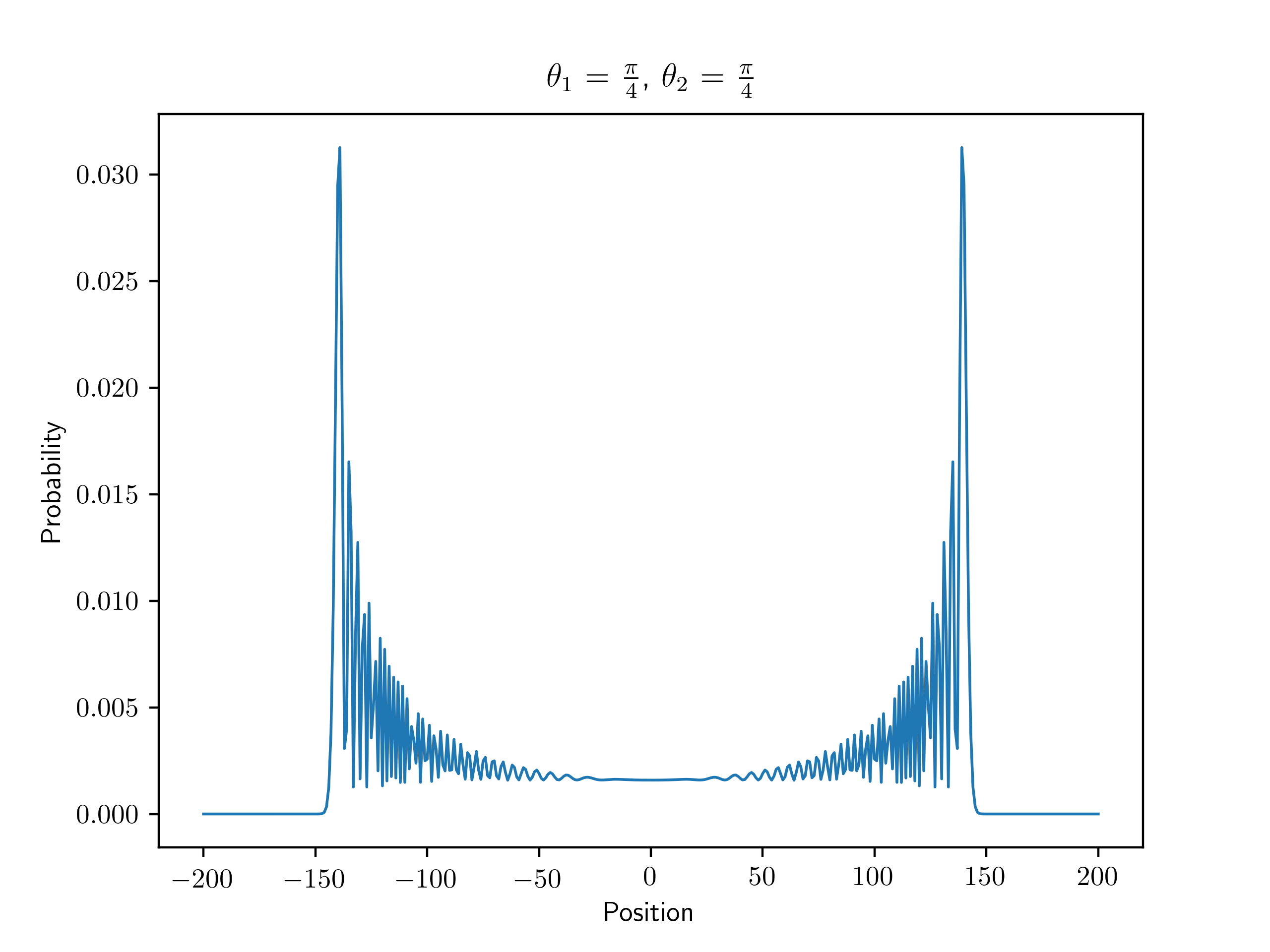} \\	
			(a) & (b)\\
			\includegraphics[width=0.45\textwidth]{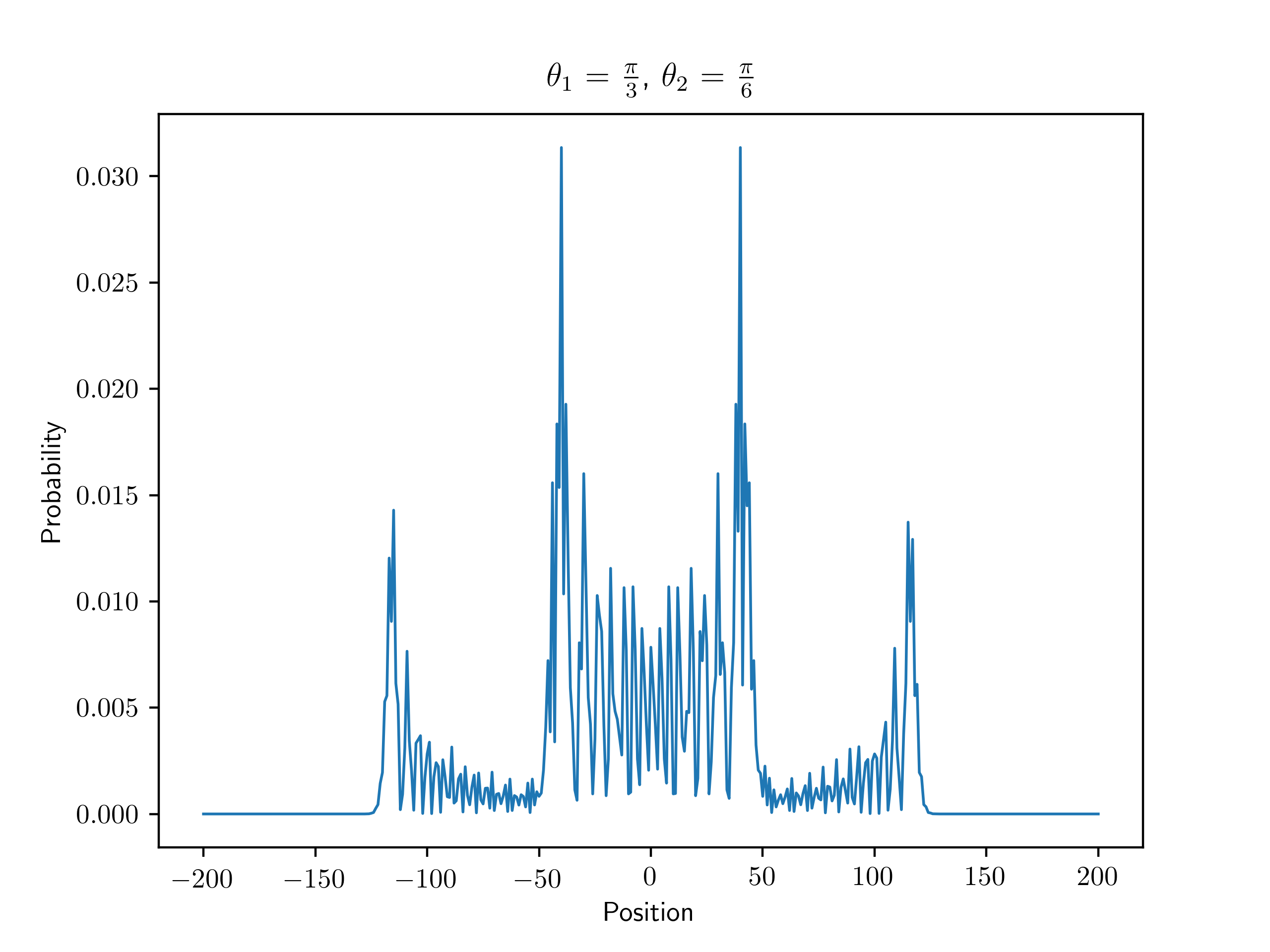} &
			\includegraphics[width=0.45\textwidth]{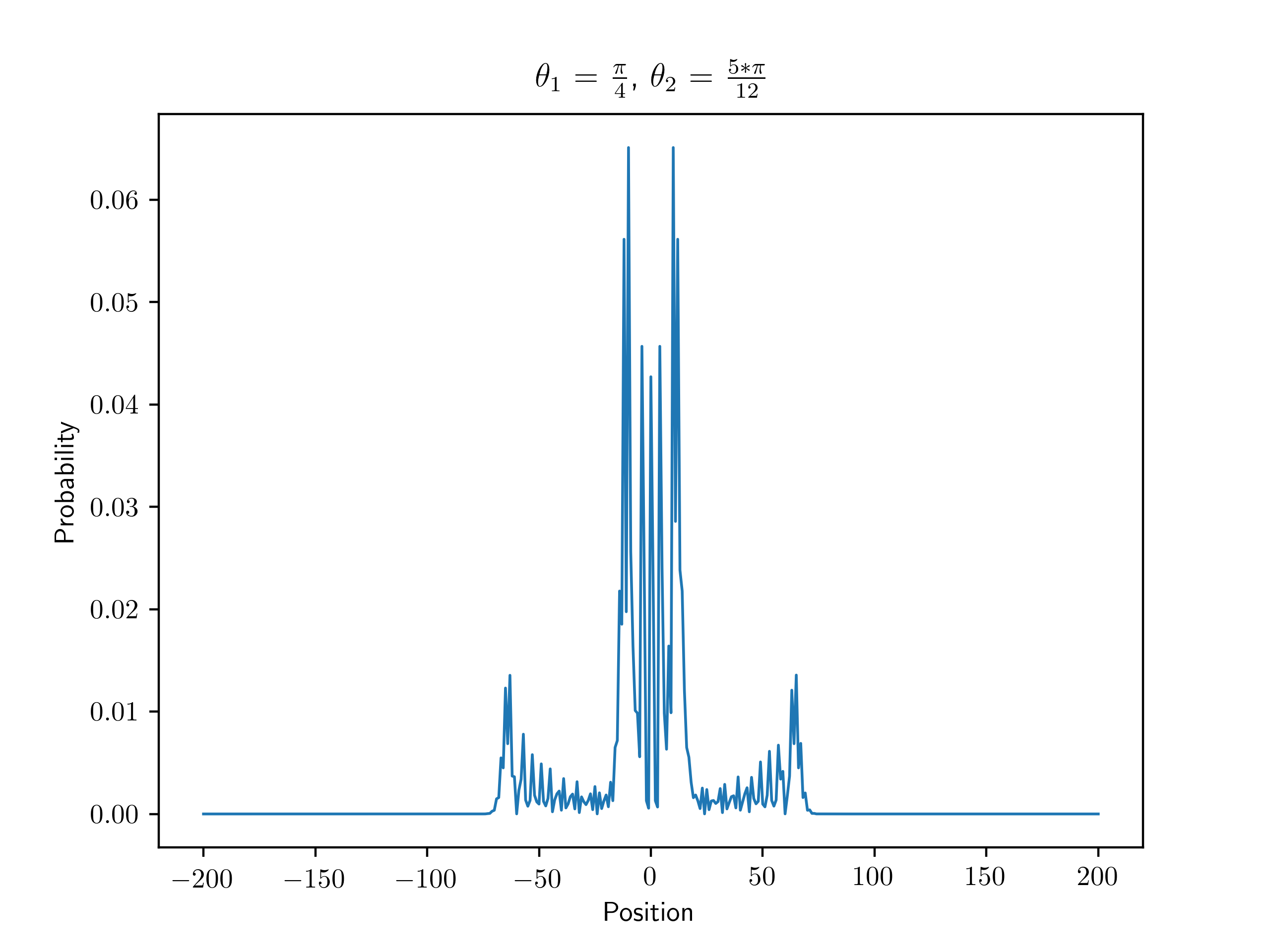} \\	
			(c) & (d)\\
		\end{tabular}
		\captionof{figure}{Position space probability distributions of a single walker performing an  SSQW in one dimension, obtained after 200 steps of walk. The figures (a), (b), (c) and (d) correspond to the cases where the parameters $\theta_1$ and $\theta_2$ have the values $(\frac{\pi}{6},\frac{\pi}{4}), (\frac{\pi}{4},\frac{\pi}{4}), (\frac{\pi}{6}, \frac{\pi}{3})$, and $(\frac{\pi}{4}, \frac{5\pi}{12})$, respectively. The distributions are symmetric as the initial state was chosen to be symmetric. The case corresponding to (b) shows an interesting characteristic of the split-step walk, namely, that the two sets of peaks coincide when the angles $\theta_1$ and $\theta_2$ are equal.
			\label{fig:fig2.2}}
	\end{minipage}
\end{table*}

%======================================================================
\section{A brief note on machine learning}
\label{sec:ml}
%======================================================================

Machine learning was originally created to try and create solutions to problems which were hard to define formally, such as recognizing faces in images, or cognition of spoken words. This necessitated the creation of algorithms that could learn from experience of examples supplied to them, so the programmer was absolved of the requirement to formally specify all knowledge needed to solve the problem beforehand. 

While the hardcoded knowledge methods worked very well for small, relatively sterile environments such as the world of chess \cite{CHH02}, such programs face difficulties in  subjective and intuitive tasks such as understanding speech or object recognition. Inference-based methods have been suggested and implemented as a possible solution to this type of problem, but have not had much success \cite{LG89, D15}. Modern approaches circumvent this difficulty by implementing algorithms that have the capability of recognising patterns in raw data on their own \cite{GBC16} - a trait now known as machine learning. 

\subsubsection*{Linear regression}

Linear regression takes an input vector of features $\mathbf{x}\in\mathbb{R}^N$ and outputs a vector $\hat{\mathbf{y}} \in \mathbb{R}^k$, which is to be interpreted as a prediction for the  actual vector $\mathbf{y} \in \mathbb{R}^k$. The model is named linear regression as it attempts to find a linear function from the input vector to the output. More mathematically, linear regression models calculate the prediction $\hat{\mathbf{y}}$ as 
\begin{equation}
	\label{eq:eq2.8}
	\hat{\mathbf{y}} = \mathbf{w}^T \mathbf{x} + \mathbf{b},
\end{equation}
\noindent
where $\mathbf{w}$ is a matrix of weights and $b$ is known as the bias vector. The weight $w_{ij}$ essentially determines how the feature $x_{i}$ affects the prediction $\hat{y}_{j}$. The bias vector is the value that this function tends to take in the absence of an input, and thus ensures that the mapping of features to predictions is an affine function.

%=================================
\subsubsection*{Ridge regression}
%====================================
Due to the large sizes of the input vectors, a case may arise where the input variables may have near-linear relationships, a phenomenon known as multicollinearity \cite{B07}. Multicollinearity leads to unbiased linear regression estimates, but with high variances. Ridge regression is a method to improve estimation by adding a small amount of bias \cite{M98}.

Assume that one requires the estimates of a vector $\mathbf{y}$, as in Eq.~(\ref{eq:eq2.8}). A traditional linear regression-based method would seek to minimize the sum of squared residuals $|| \mathbf{w}^T \mathbf{x} + \mathbf{b} - \mathbf{y} ||_2^2$, where $|| \cdot ||_2$ represents the Euclidean norm. Ridge regression introduces a regularization term in this in order to guide the solution towards preferred properties. Thus, the final minimization looks like,

\begin{equation}
	\label{eq:eq2.9}
	|| \mathbf{w}^T \mathbf{x} + \mathbf{b} - \mathbf{y} ||_2^2 + || \mathbf{\Gamma_x} ||_2^2
\end{equation}

\noindent
where $\mathbf{\Gamma_x}$ is known as the Tikhonov Matrix \cite{T63, TGSY95}. In the case when this matrix is a multiple of the identity matrix (i.e. $\mathbf{\Gamma_x} = \alpha \mathbb{I}$), this process is the same as $L_2$ regularization. In this paper, we have chosen $\alpha=0.01$. In the method of lasso regression \cite{R96}, the $L_1$ regularization is used instead.

\subsubsection*{Nearest-neighbour regression}
The $k$-Nearest Neighbour algorithm was proposed as a nonparametric method for pattern recognition, and is used for both classification and regression analyses \cite{CH67, A92}. In our case, the task is that of regression, and therefore the output is the average of k closest examples in the feature space of the training set. 
The algorithm proceeds as follows:

\begin{enumerate}
	\item Load the training data and initialize $k$ to the chosen number of neighbours.
	\item For each example in the training data,
	\begin{enumerate}
		\item Calculate the distance between the query and the current example from the data.
		\item Add the distance and index of this example to an ordered collection
	\end{enumerate}
	\item Sort the ordered collection in ascending order of distances.
	\item Pick the labels of first $k$ entries in this collection, and return the mean of the these labels.
\end{enumerate}

This algorithm defers all computation until function evaluation, and only locally approximates the said function. Thus it is also a valid example of instance-based learning \cite{PE20, H01}. As it uses the training examples in the local neighbourhood of the query to generate a prediction, it is extremely sensitive to the local structure of the data \cite{BGRS99}. Since only the $k$ closest training examples nearest to a query are considered to generate the prediction for it, this algorithm does not require an explicit training step.

\subsubsection*{Multilayer Perceptron Models}
Multilayer perceptron (MLP) models are a class of artificial neural networks, which are algorithms  modelled after biological neural networks in animal brains \cite{MP43, K56}. These algorithms learn by example, and consist of units known as neurons, which are loosely modelled after their biological counterparts. Fig.~\ref{fig:fig2.3} shows a basic MLP model with 3 neurons in the input layer, 2 hidden layers of four neurons each, and an output layer of 3 neurons. In this paper, the MLP model used has 200 neurons in the input layer, 200 in the hidden layer, and 2 in the output layer. The input and hidden layer neurons use a rectified linear unit (ReLU) as their activation function, and the output layer uses an exponential activation function. 

\begin{table*}
	\begin{minipage}[c]{\textwidth}
		~
		%\begin{figure}[!ht]
		\centering
		\begin{tabular}{cc}
			\includegraphics[width=0.45\textwidth]{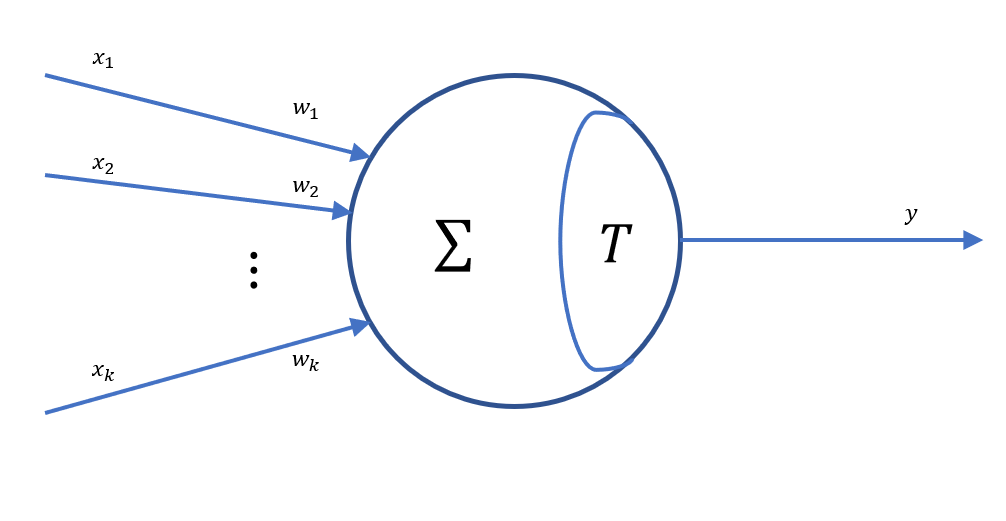} &
			\includegraphics[width=0.45\textwidth]{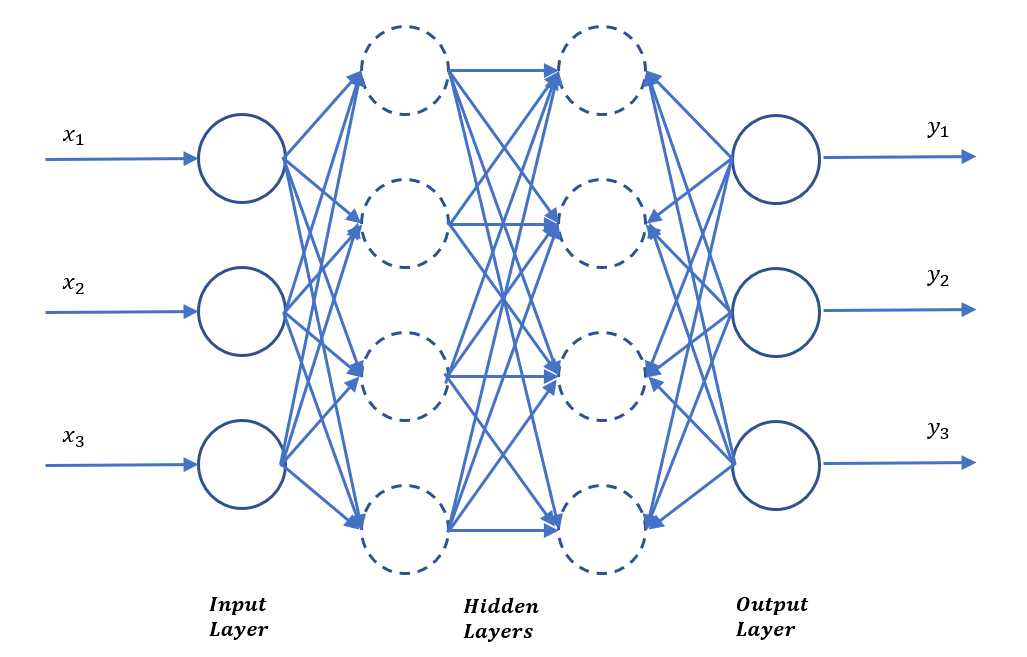} \\	
			(a) & (b)\\
		\end{tabular}
		\captionof{figure}{ A representation of a multilayer perceptron model. Figure (a) shows the structure of a single neuron, where $x_i$ are the inputs, $w_i$ are the weights, $\sum$ corresponds to the activation function, and $T$ is the threshold value for activation of this neuron. Figure (b) shows how a multitude of these perceptron units are organised in layers (thus lending the model its name), to create a simple artificial neural network. The hidden layers are drawn with dashed lines, and the input and output layers are maked in solid lines. This model has 4 neurons in each of its hidden layers, and 3 neurons each in its input and output layers. A neuron has a single available output, and each of the multiple outputs of a neuron as seen in (b) are copies of this single output value.
			\label{fig:fig2.3}}
	\end{minipage}
\end{table*}

The goal of this model is to approximate a function $\mathbf{y} = f (\mathbf{x})$ by creating a mapping $\mathbf{y} = f^* (\mathbf{x} ; \mathbf{\theta}) $ and then learning the parameters $\mathbf{\theta}$ so that $f^*$ approximates $f$ as closely as possible. This is a type of feedforward model \cite{GBC16}, as there are no inputs that are fed back to the input from the output of the model. The model is provided a certain amount of labelled examples, which consist of both the input $\mathbf{x}$ and the output $\mathbf{y}$ (also known as the {\it training set}). The model uses these to tweak the parameters $\mathbf{\theta}$ the best it can, and then predicts $\hat{\mathbf{y}}$ corresponding to unlabelled values of $\mathbf{x}$ (also known as the {\it testing set}). This technique is known as supervised learning.

Our MLP model uses the cross-entropy between training data and the predictions made by the model as the cost function, and attempts to minimize it via an optimized gradient descent method. In our model, the optimizer used is Nadam. The Nadam (Nesterov-accelerated Adaptive Moment estimation) computes adaptive learning rates for each parameter by combining the approaches of Nesterov accelerated gradients and adaptive moment estimation algorithms \cite{D16, N83, KB15, R17}.

\subsubsection*{Performance metric}

We measure the performance of our model by computing the mean square error of the model on the testing set, defined as,
\begin{equation}
	\label{eq:eq2.10}
	\text{MSE} = \frac{1}{n} \sum_{j = 1}^{k} \left[ \hat{y}_j - y_j \right]^2 ,
\end{equation}

\noindent
where $n$ is the number of elements in the testing set, and $k$ is the length of the output vector. Eq.~(\ref{eq:eq2.10}) may also be written in terms of Euclidean distance,
\begin{equation}
	\label{eq:eq2.11}
	\text{MSE} = \frac{1}{n} || \hat{\mathbf{y}}^{(test)} - \mathbf{y}^{(test)} ||_2^2
\end{equation}
of the predicted values and the actual values of $\mathbf{y}$. It is important to note here that the error is considered over the testing set and not the training set.

%======================================================================
\section{Results}
\label{sec:res}
%======================================================================

In this work, we have used machine learning algorithms and trained an MLP model in an attempt to estimate the parameters of a one-dimensional DTQW. The models were trained on DTQW with $N = 500$, and $\theta$ varying from $\frac{\pi}{180}$ to $\frac{89\pi}{180}$, at an interval of $\frac{\pi}{1800}$. The dataset consists of DTQW distributions that resulted after using these parameters in conjunction with a symmetric initial state as,
\begin{equation}
	\label{eq:eq3.1}
	\ket{\psi_0} = \frac{1}{\sqrt{2}}\left(\ket{\uparrow} + \ket{\downarrow}\right) \otimes \ket{x=0}.
\end{equation}

We have chosen the ratio of training to testing data as $3:1$ ($75\% : 25\%$) for this dataset. As a result, $668$ randomly selected probability distributions are used to train our algorithm, and the rest are used to test its performance. Three regression models, namely, K-nearest neighbours, linear regression and ridge regression, were trained on this data, and their performance was evaluated by the mean square error (MSE) of their predictions. The results are shown as in Table~\ref{tab:tab3.1}. It is observed that while all the models return a very low MSE, but the performance of the linear regression model is a few orders of magnitude better than the next best performing model (k-nearest neighbours).

\begin{table}[h]
	\begin{center}
		\begin{tabular}{|l|c|} 
			\hline
			\multicolumn{1}{|c|}{\textbf{Model}} & \textbf{Mean-squared error }\\ \hline
			Linear Regression & 2.501 x $10^{-09}$ \\ \hline 
			K-Nearest Neighbours (k = 5) & 2.058 x $10^{-06}$ \\ \hline
			Ridge Regression ($\alpha$ = 0.01) & 2.783 x $10^{-05}$ \\ \hline
		\end{tabular}
	\end{center}
\caption{Table showing MSE on test data for various machine learning models tasked with learning the coin parameter of a one-dimensional quantum walk. All the models shown here display a very low MSE, which is indicative of the fact that each of the models has successfully recognized, and is able to reproduce the patterns in the data.}
\label{tab:tab3.1}
\end{table}

Using the same training data as specified above, the model was tasked to predict the value of $\theta$, and given the testing data corresponding to $\theta=\frac{\pi}{2}$, which is outside the range used to generate the training data. Table~\ref{tab:tab3.2} details the predictions by each of these models.\\

\begin{table*}
	\begin{minipage}[c]{\textwidth}
		\begin{center}
			\begin{tabular}{|l|c|c|c|} 
				\hline
				\multicolumn{1}{|c|}{\textbf{Model}} & \textbf{Expected $\theta$} &  \textbf{Predicted $\theta$} & \textbf{Error (\%)} \\ \hline
				Linear Regression & $90^{\circ}$ & $89.956^{\circ}$ & 0.048\% \\ \hline 
				K-Nearest Neighbours (k = 5) & $90^{\circ}$ & $89.639^{\circ}$ & 0.046\% \\ \hline
				Ridge Regression ($\alpha$ = 0.01) & $90^{\circ}$ & $89.958^{\circ}$ & 0.401\% \\ \hline
			\end{tabular}
		\end{center}
		\caption{A detailing of various estimates of $\theta$ as predicted by training machine learning models. All three models predict a value of $\theta$ that is fairly close to the actual value, but the linear and ridge regression models outperform the k-nearest neighbours model in absolute value.}
		\label{tab:tab3.2}
	\end{minipage}
\end{table*}

We also implemented an MLP model (detailed in section~\ref{sec:ml}) to try and predict two parameters ($\theta$ and $N$) of a 1D-DTQW simultaneously via deep learning. The model contains $3$ layers, out of which there is $1$ hidden layer. The model uses rectified linear units as the activation function for the first two layers (i.e. the input and hidden layers, respectively), and an exponential activation for the output layer. The neural net uses the {\it Nadam} optimizer. Other possible optimizers that can be used here are AdaGrad, AdaDelta and Adam. AdaGrad offers the benefit of eliminating the need of manually tuning the learning rate, but has the weakness that it accumulates a sum of squared gradients in its denominator over time, causing the learning rate to shrink and eventually become infinitesimally small, and the algorithm stop learning at this point. AdaDelta aims to reduce the aggressive decay of the learning rate in AdaGrad by fixing a window size of the accumulated past gradients. Adam tries to improve on both AdaGrad and AdaDelta by considering an exponentially decaying average of past gradients to adjust its learning rates. Nadam, however, improves on the performance of Adam by using Nesterov-accelerated gradients, and was chosen as the best fit for this case.

To train this network, we generated a new dataset, which contained all possible combinations of the parameters $\theta$ and $N$, varying between $[\frac{\pi}{180}, \frac{\pi}{2}]$ and $[1,499]$ respectively, with a step size of $1$ for $N$, and $\frac{\pi}{180}$ for $\theta$. The dataset thus comprised of $44,910$ different probability distributions. The training-testing split was chosen to be the same as the earlier case ($75\% : 25\%$), and the training set thus consisted of a randomly selected $33,682$ probability distributions of the total. 

In order to judge the effectiveness of this model, we also designed a baseline model, which would always predict the mean of all supplied training values of $\theta$ and $N$ ($\frac{89\pi}{360}$ and $249$ respectively, for this case), for any input distribution. The MSE from this baseline model was found to be ${26077.4287}$.

Our neural network gave an output MSE of $916.0458$, which is a reduction of $96.488\%$ on the baseline error. This proves that the neural network is able to effectively learn and reproduce the patterns found in the DTQW probability distributions in order to reasonably estimate both the parameters of a 1D-DTQW.

We also tried to estimate selected parameters of an SSQW. A one-dimensional SSQW with two single-parameter coins as defined in Eq.~(\ref{eq:eq2.5}) which typically has 3 parameters, $\theta_1$, $\theta_2$, and $N$. For the purposes of this task, we varied $\theta_1$ in the range $[\frac{\pi}{1800}, \frac{\pi}{2}]$ at an interval of $\frac{0.04\pi}{180}$, while keeping $\theta_2 = \frac{\pi}{4}$, and $N=100$ fixed. We trained a linear regression model on this data, as well as a K-nearest neighbour model. The parameter was estimated with a good accuracy, as is seen from the contents of Table~\ref{tab:tab3.3}.

\begin{table}[h]
	\begin{center}
		\begin{tabular}{|l|c|} 
			\hline
			\multicolumn{1}{|c|}{\textbf{Model}} & \textbf{Mean-squared error}\\ \hline
			K-Nearest Neighbour (k = 5) & $3.112 \times 10^{-07}$ \\ \hline
			Linear Regression & $1.265 \times 10^{-08}$ \\ \hline 
		\end{tabular}
	\end{center}
\caption{Mean squared error on test data for ML models in predicting a single parameter in a split-step quantum walk. Both the models are able to learn the parameters well, but it can be seen that the linear regression model still yields an MSE that is an order of magnitude lower than the K-nearest neighbours model.}
\label{tab:tab3.3}
\end{table}

%======================================================================
\section{Conclusions}
\label{sec:conc}
%======================================================================
In this work, we have demonstrated the effectiveness of machine learning models while estimating the coin parameter in a DTQW and an SSQW. We have applied these models on the one-dimensional DTQW and SSQW, and attempted to estimate the parameters from these walks.

In the case of DTQW, we applied three different models in order to estimate the coin parameter $\theta$, and conclude that a machine learning-based approach is indeed able to estimate the parameter very well. We also attempt to use a neural network in order to try and predict the two parameters at once, and show that its prediction error reduces by a significant amount from the baseline error, implying that the network is able to distinguish and replicate patterns found in this data. We also use two different models in order to estimate one of the coin parameters of a SSQW.

It is important to keep in mind that the accuracy of the predictions will vary with the amount of training data it is given. Typically, larger datasets improve the accuracy. The accuracy is also very dependent on the model itself under consideration. There is no single known machine learning algorithm that may outperform all others, and it is thus important to choose the algorithm for implementation with care.

\vskip 0.2in
\noindent
{\bf Acknowledgment:} 
CMC would like to thank Department of Science and Technology, Government of India for the Ramanujan Fellowship grant No.:SB/S2/RJN-192/2014. We also acknowledge the support from Interdisciplinary Cyber Physical Systems (ICPS) programme of the Department of Science and Technology, India, Grant No.:DST/ICPS/QuST/Theme-1/2019/1 and US Army ITC-PAC contract no. FA520919PA139.

%======================================================================

%======================================================================

\end{document}